# Multi-channel frequency router based on valley-Hall metacrystals

*Jiayu Fan[1], Haitao Li[1], Shijie Kang[1,2], Peng Chen[1,3], Biye Xie[4], Fang Ling[1,5], Ruping Deng[6], Xiaoxiao Wu[1,2, *]*

[1]Modern Matter Laboratory and Advanced Materials Thrust, The Hong Kong University of Science and Technology (Guangzhou), Nansha, Guangzhou 511400, Guangdong, China

[2]Low Altitude Systems and Economy Research Institute, The Hong Kong University of Science and Technology (Guangzhou), Nansha, Guangzhou 511400, Guangdong, China

[3]Quantum Science Center of Guangdong-HongKong-Macao Greater Bay Area, Shenzhen-Hong Kong International Science and Technology Park, Futian, Shenzhen 518000, Guangdong, China

[4]School of Science and Engineering, The Chinese University of Hong Kong, Shenzhen, Guangdong, 518172, China

[5]School of Physics and Electronic Engineering, Sichuan University of Science & Engineering, Zigong, 643000, China

[6]Institute of Modern Optics, Nankai University, Tianjin 300350, China

[*] To whom correspondence should be addressed. E-mail: xiaoxiaowu@hkust-gz.edu.cn

**Abstract:**
Topological photonics has revolutionized manipulations of electromagnetic waves by leveraging various topological phases proposed originally in condensed matters, leading to robust and error-immune signal processing. Despite considerable efforts, a critical challenge remains in devising frequency routers operating at a broadband frequency range with limited crosstalk. Previous designs usually relied on fine tuning of parameters and are difficult to be integrated efficiently and compactly. Here, targeting the demand for frequency-selective applications in on-chip photonics, we explore a topological approach to photonic frequency router via valley-Hall metacrystals. Diverging from the majority of studies which focuses on zigzag interfaces, our research shifts the attention to armchair interfaces within an ABA sandwich-like structure, where a single column of type-B metacrystal acts as a perturbation in the background type-A metacrystal. Essentially, through tuning a single geometric parameter of the type-B metacrystal, this configuration gives rise to interface states within a customized frequency band, enabling signal routing with limited crosstalk to meet specified demands. Moreover, this concept is practically demonstrated through a photonic frequency router with three distinct channels, experimentally exhibiting robust wave transmissions with excellent agreement with the design. This investigation manifests possible applications of the armchair interfaces in valley-Hall photonic systems and advances development of photonic devices that are both compact and efficient. Notably, the approach is naturally compatible with on-chip photonics and integration, which could benefit telecommunications and optical computing applications.

Drawing inspiration from quantum Hall effect[1-8], quantum spin Hall effect[9-15], and quantum valley Hall effect[16-20], researcher has explored photonic topological phenomena which opens new avenues for manipulating light in unprecedented ways[1, 21-24]. In a large number of topological effects, photonic valley-Hall effect has emerged as a particularly intriguing concept, drawing intense research interest due to its potential for robust, backscattering-immune waveguiding. The valley-Hall effect, akin to its electronic counterpart, leverages the valley degree of freedom in photonic crystals to control the flow of light[25-32]. To date, the majority of studies have focused on zigzag interfaces of valley-Hall photonic crystals, where the valley-Hall interface states span nearly entire bandgap[13, 19, 20, 25, 32, 33], offering broad operational bandwidth for photonic devices. However, this near traverse of bandgap naturally impedes the design of photonic devices with different frequency channels for frequency multiplex functions. In practical scenarios, frequency multiplex is important for many applications, such as photonic computing, photonic communication, and photonic interconnection. Obviously, more channels a frequency multiplex device have, the larger information capacity the device has.

Previously, some frequency routers have been proposed[34, 35]. A dual-band frequency router based on topological metamaterial is proposed[34], which employs zigzag interface states. Another research group has proposed a frequency router based on topological metamaterial, which also achieved the function of dual-channel topological frequency router[35]. Despite these advancements, creating a multi-channel frequency router that allows each channel to function independently with little cross talk is still a challenge due to the limitation imposed by the characteristics of the zigzag interface.

To address this urgent issue, in this work, we shift our attention to the much lesser-explored armchair interfaces in a sandwich structure configuration. This configuration consists of a single column of type-B valley-Hall metacrystal flanked by type-A valley-Hall metacrystals on either side. Interestingly, this arrangement facilitates the emergence of interface states that occupy a relatively narrow frequency range, a stark contrast to the broad bandwidth typically associated with zigzag interfaces. The spectral extent of these interface states is found to be highly sensitive to geometric parameters of type-B valley-Hall metacrystal.

Leveraging the tunable frequency range of these interface states, we propose a design for a photonic frequency router. The router exploits the selective frequency guidance at the armchair interfaces to achieve precise control over the propagation of different frequency components, a critical function for advanced photonic circuitry. The potential applications of such a frequency router are vast and could revolutionize fields ranging from telecommunications to optical computing. This study not only expands the understanding of valley-Hall metacrystals but also underscores the practical implications of topological photonic states in the development of functional, miniaturized, and highly efficient photonic devices. We believe that our findings pave the way for robust and versatile on-chip light manipulation, holding great promise for the future of photonic technology.

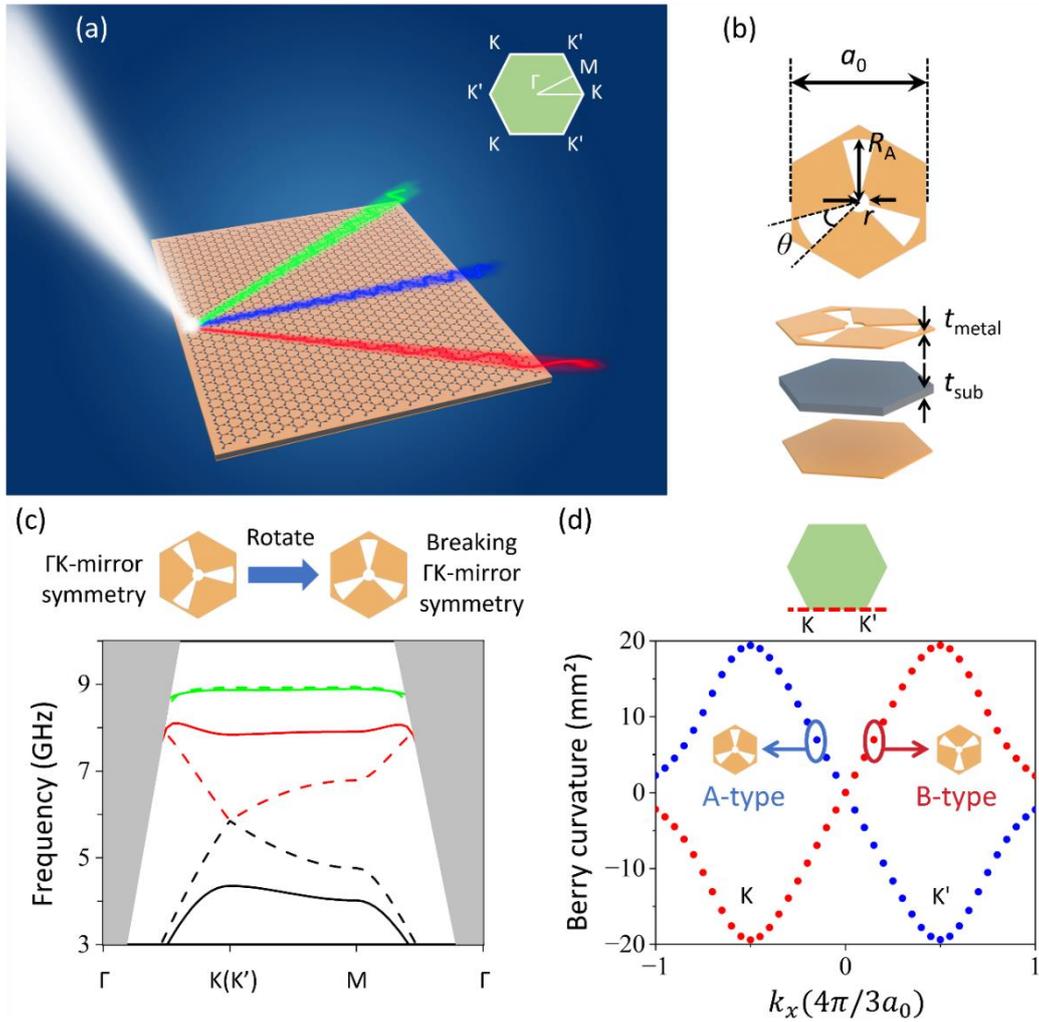

**Figure 1**. (a) Conceptual representation of an integrated on-chip frequency router with three distinct propagation channels. Electromagnetic waves of varying frequencies enter the metamaterial and follow predetermined paths accordingly. (b) Detailed three-layer construction of a single unit cell, with an overhead perspective illustrating the topmost layer's configuration. (c) Comparative band structures of the unit cell, delineating the pre-symmetry (dotted line) and post-symmetry-breaking (solid line) states. The region out of light cone is covered by gray region. (d) Berry curvature (*z*-component) of the 1st band for A- and B- type structures along $k_x$ when $k_y = -2\pi/\sqrt{3}\,a_0$, directly calculated from COMSOL Multiphysics.

A conceptual design of a frequency-selective routing device with three distinct pathways is depicted in Figure 1(a). EM waves of varying frequencies are guided along specific pathway, represented by three unique colors correlating to different frequencies. The structure is configured as a triangular-lattice crystal, characterized by a lattice constant *a* = 12 mm. An inset within Figure 1(a) presents the first Brillouin zone for clarity. A single unit cell is elaborated upon in Figure 1(b), composed of a sandwich-like structure: two metallic layers flanking a central dielectric spacer. The top and bottom metallic layers are covered on a dielectric layer of the permittivity 3.48 and thickness $t_{sub}$ = 0.762 mm. A top-down view in Figure 1(b) illustrates the geometric configuration of upper metallic layer. The thickness of top and bottom metallic layers are both $t_{metal}$ = 35 μm. The top metallic layer features a slot pattern, comprising an inner circle with a radius *r* = 1 mm and three outward-extending radial arms, each arm with a sector angle of *θ* = 30° and a fan radius of $R_A$ = 6 mm. We plot the band structure of a ΓK-mirror symmetric unit cell in Figure 1(c) with the dashed lines. It is clear to see that the 1st band and the 2nd band linearly cross each other at the K and K' valleys to form degenerate Dirac points. By rotating the structure 30° to break the ΓK-mirror symmetry, a wide band gap is created, providing the prerequisite of frequency router. The band structure of the unit cell, which we refer to as A-type, is represented by the solid line in Figure 1(c). Furthermore, we analyze the dependent of bandgap of the unit cell after breaking ΓK mirror symmetry (refer to Supplementary Note 1 in Supplementary Information (SI)). In addition, a parameter *α* is considered as a perturbation to introduce break of ΓK-mirror symmetry of unit cell, which lifts the degeneracy at valleys and hence open a bandgap (refer to Supplementary Note 2 in SI). As the electric filed maps display in Figure S5 (b) and (c), the energy flux of the two states is either left circularly polarized or right circularly polarized. In order to analyze and demonstrate the difference of valley-Hall topological phase, we calculate the Berry curvature (BC) of the structure along $k_x$ when $k_y = 2\pi/\sqrt{3}a_0$ after breaking for A-type structure. Where the BC is defined by

$$\mathbf{F}_\mu(\mathbf{k}) = \nabla_k \times i\langle \mathbf{E}_\mu(\mathbf{k}) | \nabla_k \mathbf{E}_\mu(\mathbf{k}) \rangle \tag{1}$$

The BC can be directly calculated with the finite-element simulations[33], and its distribution, only the non-zero *z*-component, is shown in Figure 1(d) as the blue dotted line, which verifies that non-trivial localized BC emerges for A-type structure. On the other hand, we rotate the A-type structure by 180°, leading to a structure we refer to as B-type, and also calculate its BC, shown as red dotted line in Figure 1(d). It is obvious that they have localized but exactly opposite BCs, which indicates that A- and B-type structures are of opposite valley-Hall effects. Thus, what will happen if we "dope" some B-type structures into A-type structures?

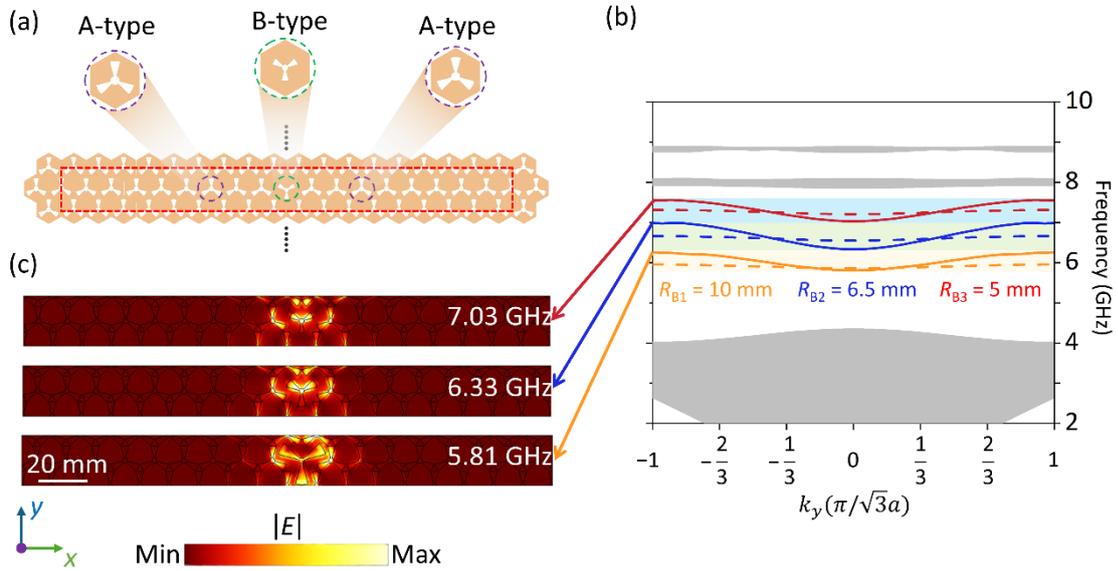

**Figure 2.** (a) Diagram of supercell. (b) Band structure of super cell when $R_{B1}$ = 10 mm, $R_{B2}$ = 6.5 mm and $R_{B3}$ = 5 mm and (c) the corresponding Electric field maps of the super cell when $k_y = 0$.

Therefore, we consider a "minimal" case, that is, we consider a sandwich-like supercell, but with only one B-type structure in the supercell. We systematically arrange the A- and B-type structures into a sandwich-like sequence, we create a supercell as depicted in Figure 2(a). The A-type structure demarcated by a purple dotted line and the B-type structure emerges through rotation of the A-type, highlighted by a green dotted line. We note that our strategy is to arrange the unit cells to form an armchair interface. The key reason is that the valley-polarized interface states from a zigzag interface will nearly traverse the band gap which makes it nearly impossible to design a frequency router. Further, we find that the variation of the fan dimension $R_B$ of the B-type structure resulting interface states at different frequencies, allowing for the tailored control of the frequency by fine-tuning the B-type's size which enables frequency-selective routing. To demonstrate the potential of this variation, we choose three cases, that is, the radius of B-type structure is selected as $R_{B1} = 10$ mm, $R_{B2} = 6.5$ mm, and $R_{B3} = 5$ mm, respectively. Three pairs of interfaces states emerge in the bandgap of A-type metacrystal, colored in orange, blue, and red in Figure 2(b), with the bulk states represented as gray shaded regions. It is noteworthy that the chosen B-type structures resulting in the interface states exhibiting minimal overlap with one another and bulk states. Obviously, this separation of interfaces states allows for three different frequency channels. Each case manifests two interface states corresponding to symmetric and anti-symmetric modes. The dotted lines represent the anti-symmetric modes' dispersions, while the solid lines correspond to the symmetry modes, as detailed in electric field maps shown in Supplementary Note 3 (refer to SI). The electric field maps of the symmetric modes when $k_y = 0$ at different frequencies are displayed in Figure 2(c), where a pronounced interface state is evident. It is crucial to emphasize that even with the incorporation only a single B-type structure within the supercell, the interface states are tightly confined around the interface thanks to the wide bandgap of A-type metacrystal. Consequently, we can design a crosstalk-free frequency router with three distinct channels by judiciously arranging A- and B-type structures. The dependence of the frequency range of the interface states on the radius $R_B$ is summarized in Supplementary Note 4 (see SI). Furthermore, in selecting the interface orientation for our supercell configuration, we carefully considered the implications for our frequency router design. An armchair interface was ultimately chosen because its band structure more effectively met our design requirements, in contrast to the zigzag interface, which was found to be less suitable, as detailed in Supplementary Note 5 of SI. Supplementary Note 6 explores armchair interface configurations, discussing how to achieve flatter band structures which is crucial for our design.

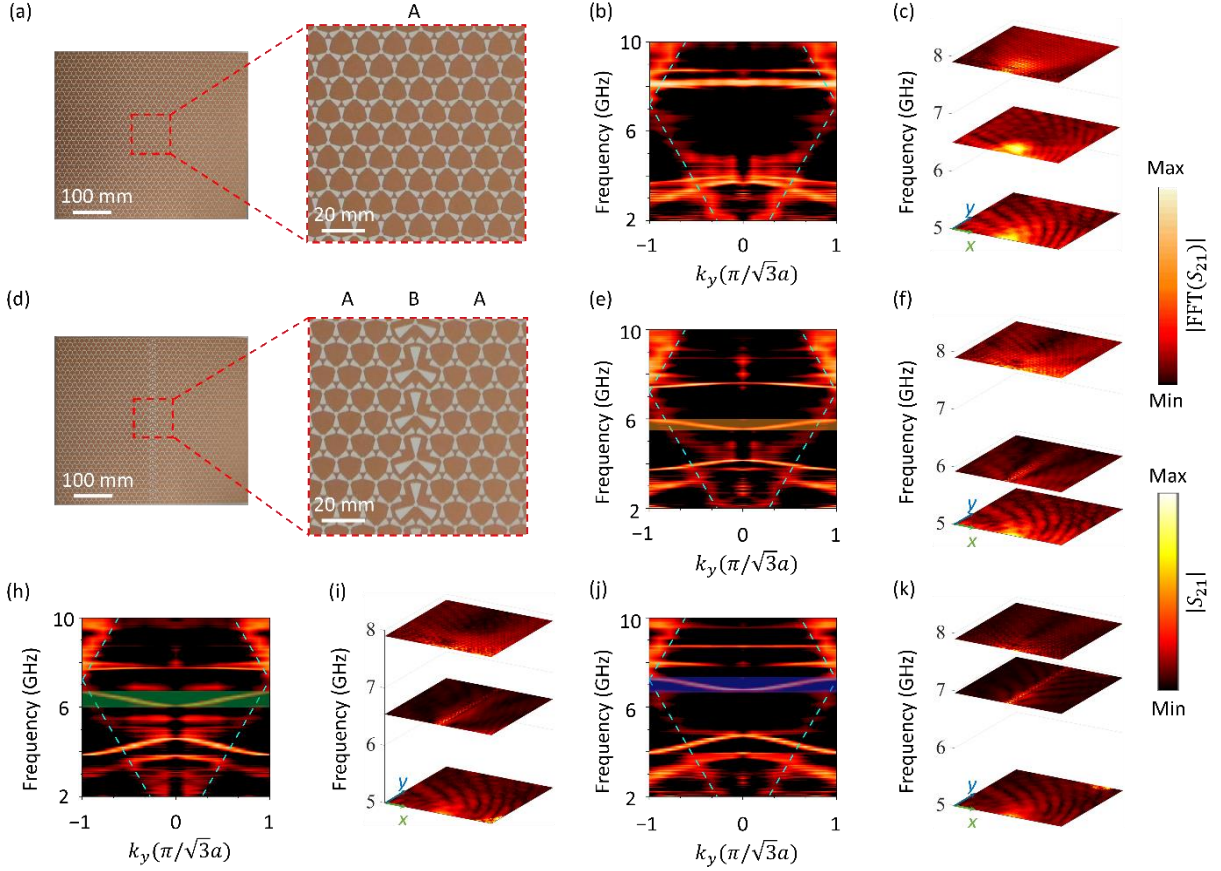

**Figure 3.** Experimental result of supercell. (a) The supercell sample consists of A type structure with (b) corresponding experimental result for band structure and (c) electric field maps at different frequencies. (d) The supercell sample incorporating a single B type structure with the fan size of $R_{B1}$ = 10 mm. The experimental results also cover the band structure of ABA type sample with the radius of A type structure ($R_A$ = 6 mm) and incorporating a single B type structure with the fan sizes of (e) $R_{B1}$ = 10 mm, (h) $R_{B2}$ = 6.5 mm and (j) $R_{B3}$ = 5 mm. And the corresponding electric field maps at different frequencies when (f) $R_{B1}$ = 10 mm, (i) $R_{B2}$ = 6.5 mm and (k) $R_{B3}$ = 5 mm.

Then we proceed to verify our design. We first demonstrate a sample shown in Figure 3(a) which consists of A-type structure for verifying the band structure. A near field scanning platform is employed to obtain the electric field maps 1 mm above the samples. Then we can obtain the experimental band structure of the samples by two-dimensional fast Fourier transform (2D FFT). Figure 3(b) shows the experimental band structure of supercell consisting of only A type structure. A clear bulk state below 4 GHz and near 8 GHz can be observed, which can agree well with the simulated result shown in Figure 2(b). The outline of light line (folded) is represented by the cyan dashed lines, which are also band folded. The three electric field maps at 5GHz, 6.5GHz and 7.9GHz are shown in Figure 3(c). What we observe are generally EM waves in the free space, as manifested by bright strips within the light cone. To verify the existence of interface state, we then consider the sample of supercell incorporating a single B type structure with fan size of $R_{B2}$ = 10 mm shown in Figure 3(d). As shown in Figure 3(e), an interface state shaded in orange is observed clearly, and it agrees well with the simulated result in Figure 2(b). Figure 3(f) shows the electrical field maps at different frequencies. When $f$ = 5 GHz, a frequency in the bandgap, no interface state exists and what we observe are generally EM waves in free space. For higher frequency $f$ = 5.92GHz, the EM wave propagates along the middle path as we design, verifying that there is an interface state indeed. When the frequency further increases to $f$ = 7.9 GHz, what we observe is again the EM waves in the free space. When we decrease the radius of B-type structure to $R_{B2}$ = 6.5 mm and $R_{B3}$ = 5 mm, respectively, clear interface states are also observed in the band structures as shown in Figure 3(h) and (j). The electrical field maps in Figures 3(i) and 3(k) also confirm the presence of interface states. The slight frequency shift in band structure compared with simulations can be attributed to the fabrication tolerance, such as variations of dielectric permittivity. According to the simulation results depicted in Figure 2(b), each interface state is expected to have two modes: one symmetric mode and one anti-symmetric mode. However, in the experimental data, we only observe one interface mode for each band. This discrepancy arises because the frequency range of the symmetric modes overlaps with that of the anti-symmetric modes, which is further discussed in Supplementary Note 7 of SI. The overlap between these modes results in the experimental concealment of anti-symmetric modes by more dominant symmetric modes. Moreover, additional interface states at lower frequencies are present as indicated in Figure 3(e), (h), and (j). Despite their presence, these lower frequency interface states are not considered in our simulation analysis. As detailed in Supplementary Note 8 in the SI, the reason for their exclusion is due to the lack of precise control over these states, which is a critical factor for the design of a frequency router.

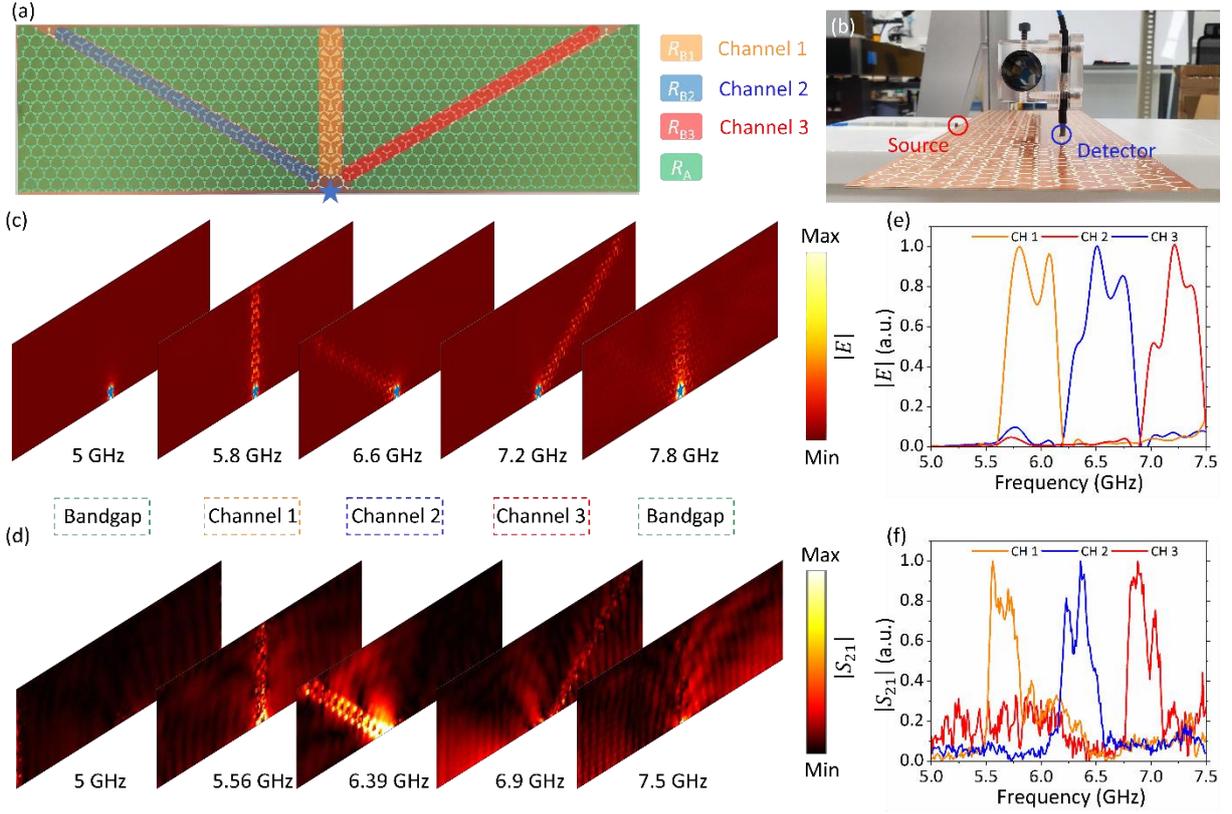

**Figure 4.** (a) Photograph of experimental setup. Two monopole antennas are used in experiments (see Methods section). Frequency router is excited by one of the antennas and in plane component of transmitted wave is measured by the other one. Where $R_{B1} = 10$ mm, $R_{B2} = 6.5$ mm and $R_{B3} = 5$ mm. (b) Top view photograph of the frequency router with three channels working on different frequency region. (c, d) Field maps from simulations (c) and experiments (d) for the frequency router excited at corresponding frequencies. Normalized spectra of electric field amplitude ($|S21|$ in experiments) at output terminals of three channels retrieved from simulations (e) and experiments (f).

After we have verified that the interface states reside at different frequency intervals, we now design a sample to function as the frequency router. Figure 4(a) shows the experimental setup, including source and detector. Figure 4(b) depicts the sample with three channels operating at different frequencies. In experiments, the linear-polarized source is attached at the position indicated by blue star and a near-field electric probe is used to measure electric field point-by-point above the sample. Figure 4(c) shows the simulated electrical field maps 0.5 mm above the frequency router. No interface state or bulk state is transmitted on the devices when the frequency is 5 GHz as it falls within band gap. As the frequency increases from 5 GHz to 5.8 GHz, the first channel starts operating, and an interface state can be observed, which agrees well with the band structure shown in Figure 2(b). We can also observe the other two channels as the frequency increases to 6.6 GHz and 7.2 GHz. When frequency increases to 7.9 GHz, we observe that some modes can propagate on the devices as they are outside band gap region. It is worth noting that these three channels can be designed to propagate in different paths, as a single column can guide the EM wave on frequency router, and the working frequency region have no overlap with each other or bulk state. Figure 4 (d) shows the experimental field maps of the frequency router, 1 mm above the sample, which agrees well with the simulated results. Additionally, we retrieved and summarized the normalized spectra of electric field amplitude at output terminals of three distinct channels from simulated and experimental results, as shown in Figures 4(e) and 4(f), respectively. The frequency shifts are mainly caused by the inaccuracies in permittivity of dielectric layer.

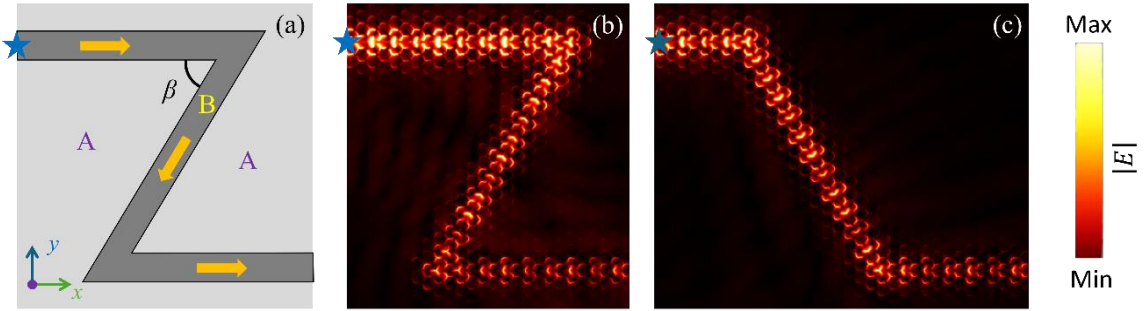

**Figure 5.** Robustness of interface state with different corner angle when $R_{B2} = 6.5$ mm. (a) Schematic diagram of a Z-Shape topological interface state within a sandwich-like path formed by A- and B- type structures. The corner angel is $\beta = 60°$. Simulated field map angle at 6.86 GHz when corner angle is (b) 60° and (c) 120°.

We further analyze the robustness of metacrystal device and designed a waveguide with different bending angles. Initially, we considered a Z-shaped waveguide featuring two sharp corners as shown in Figure 5(a), in which the corner angle $\beta = 60°$ and radius of B type structure is $R_{B2} = 6.5$ mm. The waveguide is essentially a curved ABA-type domain. The simulated result is shown in Figure 5(b), in which the blue star represents the source. Despite the presence of little scattering losses at the bends, an effective interface state is observed along the path. We then conducted a numerical investigation with a geometric configuration featuring a corner angle $\beta = 120°$, and again, an effective interface state is also observed along the path as shown in Figure 5(c). The overall results confirm that the bends only have little influence on transport of interface state.

In conclusion, we have demonstrated a frequency-selective routing mechanism using valley-hall metacrystals with minimal crosstalk. A single B-type structure column acts as a perturbation in an A-type dominated supercell, guiding EM waves along specific paths. This approach allows for adjusting the operating frequency by simply changing the perturbation size, showcasing the router's customizability. Our research also underscores the potential of the armchair interface state. Experimental results reveal three independent channels that match simulated outcomes, with slight frequency shifts due to fabrication and material flaws. However, the robustness of the interface states induced by valley-Hall effect ensures reliable wave transmission, even with waveguide bends. Therefore, our work presents a systematic design approach for multi-channel frequency routers that is easy for integration, which could enhance performance and reliability of on-chip photonic devices.

**ACKNOWLEDGEMENT**

This research was supported by the National Natural Science Foundation of China (No. 12304348), the support from the Quantum Science Center of Guangdong-Hong Kong-Macao Greater Bay Area (Guangdong), Guangzhou Municipal Science and Technology Project (No. 2023A03J0003), Guangzhou Municipal Science and Technology Project (No. 2024A04J4351), Research on HKUST(GZ) Practices (HKUST(GZ)-ROP2023021). Dr. Xie acknowledges support from the National Key R&D Program of China (2023YFA1407700), Stable Support Program for Higher Education Institutions of Shenzhen (No. 20220817185604001), Guangdong Basic and Applied Basic Research Foundation (No. 2024A1515012031). Dr. Chen acknowledges the Guangdong Provincial Quantum Science Strategic Initiative (No. GDZX2302005). Dr. Ling acknowledges support from Scientific Research and Innovation Team Program of Sichuan University of Science and Engineering (SUSE652B004).

**DATA AVAILABILITY**

The data that support the findings of this study are available from the corresponding author upon reasonable request.

**SUPPLEMENTAL MATERIAL**

See the Supplemental Material for detailed analyses on the influence of geometric modifications and material properties on the bandgap and interface states (including the effects of fan radius, circle radius, fan angle, permittivity, and thickness of dielectric layers on the bandgap), investigations into circularly polarized states at valleys and their perturbations, the role of supercell configurations on interface states, and the dependence of interface state frequency ranges on structural dimensions.